# The photometric structure of the inner Galaxy

James Binney[1] Ortwin Gerhard[2] and David Spergel[3]
[1] *Theoretical Physics, Keble Road, Oxford OX1 3NP*
[2] *Astronomisches Institut, Universität Basel, Venusstrasse 7, CH-4102 Binningen, Switzerland*
[3] *Peyton Hall, Princeton University, Princeton, NJ 08544, U.S.A.*

**ABSTRACT**
The light distribution in the inner few kiloparsecs of the Milky Way is recovered non-parametrically from a dust-corrected near-infrared COBE/DIRBE surface brightness map of the inner Galaxy. The best fits to the photometry are obtained when the Sun is assumed to lie $\sim 14 \pm 4\,\mathrm{pc}$ below the plane. The recovered density distributions clearly show an elongated three-dimensional bulge set in a highly non-axisymmetric disk. In the favoured models, the bulge has axis ratios 1:0.6:0.4 and semi-major axis length $\sim 2\,\mathrm{kpc}$. Its nearer long axis lies in the first quadrant. The bulge is surrounded by an elliptical disk that extends to $\sim 2\,\mathrm{kpc}$ on the minor axis and $\sim 3.5\,\mathrm{kpc}$ on the major axis.

In all models there is a local density minimum $\sim 2.2\,\mathrm{kpc}$ down the minor axis. The subsequent maximum $\sim 3\,\mathrm{kpc}$ down the minor axis (corresponding to $l \simeq -22°$ and $l \simeq 17°$) may be associated with the Lagrange point $L_4$. From this identification and the length of the bulge-bar, we infer a pattern speed $\Omega_b \simeq 60 - 70\,\mathrm{km\,s^{-1}\,kpc^{-1}}$ for the bar.

Experiments in which pseudo-data derived from models with spiral structure were deprojected under the assumption that the Galaxy is either eight-fold or four-fold symmetric, indicate that the highly non-axisymmetric disks recovered from the COBE data could reflect spiral structure within the Milky Way if that structure involves density contrasts greater than $\gtrsim 3$ at NIR wavelengths. These experiments indicate that the angle $\phi_0$ between the Sun–centre line and a major axis of the bulge lies near $20°$.

## 1 INTRODUCTION

There is now compelling evidence from several different sources that the centre of the Milky Way is non-axisymmetric – see Blitz & Teuben (1996) for reviews. Since the inner Galaxy's projected brightness distribution is a function of the two coordinates of the sky, while a non-axisymmetric luminosity density is a function of all three spatial coordinates, it would be natural to assume that the latter cannot be unambiguously determined from the former. Binney & Gerhard (1996; hereafter Paper I) show, however, that it is possible in many cases to recover accurately the three-dimensional luminosity distribution of a model Galaxy from its projected density provided the model has three mutually orthogonal mirror planes of known orientation. Thus in the case of eight-fold symmetry the ambiguity inherent in the determination of the bulge luminosity density can be largely reduced to that involved in choosing the orientation of the bulge's mirror planes. Moreover, in Paper I it was found that one can obtain valuable guidance as to the orientation of these planes. Here we report the results of applying this approach to surface-brightness distributions that are based on those measured by the DIRBE experiment aboard the COBE satellite.

COBE/DIRBE measured the sky brightness in both the near- and the far-infrared. The near-IR luminosity derives almost entirely from stars, and thus provides a powerful probe of the density of stars within the Galaxy. Unfortunately, even at wavelengths of a few microns there is non-negligible extinction along lines of sight towards the Centre, and it is essential that corrections for this extinction be made before the data are used to infer the Galaxy's luminosity density. Arendt et al. (1994) estimated the extinction along different lines of sight by assuming that the bulge has a constant $J - K$ colour and is obscured by a foreground screen of dust. Spergel, Malhotra & Blitz (1996; hereafter Paper II) showed that a significantly more reliable estimate of the effects of extinction can be obtained by modelling the observed luminosity at $\sim 240\,\mu\mathrm{m}$. This is dominated by thermal emission from dust, so with some additional assumptions one can deduce from it the spatial distribution of dust. The extinction model derived in this way in Paper

II is fully three-dimensional rather than constituting a foreground screen. It is, moreover, non-axisymmetric and incorporates a radial gradient in the temperature of the dust. Its reliability is attested by its ability to predict, from the far-IR data, $J - K$ reddening indices at each point on the sky that agree well with those inferred from the near-IR data under the assumption of a constant intrinsic colour within the bulge. Thus this extinction model constitutes a material advance on that of Arendt et al..

Since they assumed that the obscuring dust forms a foreground screen, Arendt et al. could correct the near-IR measurements for the effects of extinction and thus obtain the surface-brightness distribution that would be observed in the absence of dust. The three-dimensional dust model of Spergel et al. does not permit an unambiguous correction of the data for obscuration – the amount by which radiation that reaches us from any direction has been dimmed depends upon the point at which it was emitted and therefore on the adopted luminosity model. However, by fitting an axisymmetric model to the data one can make a reasonable estimate of the true dimming and use this estimate to correct the data for obscuration. We use the COBE data as corrected by Spergel et al. in this way.

Dwek et al. (1995) fitted the corrected data of Arendt et al. to parameterized functional forms of the bulge's luminosity density. The analysis below differs from that of Dwek et al. in that (i) we use corrected data of Spergel et al., and (ii) we infer the bulge's luminosity density by the deprojection technique of Paper I. We concentrate on results obtained from the data for the $L$ and $M$ bands (centred on $3.5\,\mu\text{m}$ and $4.9\,\mu\text{m}$, respectively), since these have the smallest extinction corrections.

The paper is organized as follows. In Section 2 we describe modifications to the technique described in Paper I that we have found essential to cope with the high level of noise in the COBE data. In Section 3 we describe the fits we have obtained to the photometry, and the three-dimensional luminosity distributions that underlie them. In Section 4 we discuss the implications of spiral features for our analysis. Section 5 sums up and assesses the physical plausibility and possible dynamical interpretations of various models. We assume throughout the Sun lies 8 kpc from the Galactic centre.

## 2 NOISE AND SMOOTHING

Paper I described a Richardson–Lucy algorithm for the deprojection of surface photometry. Tests in Paper I of the algorithm's ability correctly to deproject measured surface-brightness distributions assumed that the dominant source of noise within the data would be Poisson fluctuations in the stellar number density. In this case, our data, which have a fixed angular resolution $1.5°$, would be relatively noise-free near the centre. Unfortunately, on applying our algorithm to the cleaned data of Paper II, it became clear that the central data are more noisy than this assumption predicts, presumably because, as Kent, Dame & Fazio (1991) concluded, residual small-scale fluctuations in the extinction are the dominant source of noise in near-infrared photometry of the Milky Way. Presented with unexpectedly noisy data, the algorithm of Paper I generated implausibly noisy three-dimensional luminosity densities.

We experimented with two possible remedies for this unsatisfactory situation. The first was Lucy's (1994) modification of the standard Richardson–Lucy algorithm. This involves introducing a 'prior', that is, a luminosity density to which one assigns high a priori probability, and determining at each stage in the iterations the entropy of the current model with respect to this prior. Lucy modifies the iterative steps in such a way that the final model maximizes a linear combination of the likelihood and the entropy. The entropy's weight in this linear combination is a parameter $\alpha$: when $\alpha$ is very large, the recovered distribution differs little from the assumed prior, while for $\alpha = 0$ the original Richardson–Lucy algorithm is effectively recovered. Unfortunately, we did not find an interesting range of values of $\alpha$ within which the recovered distribution was both smooth and differed significantly from the assumed prior.

For this reason our results rely on an alternative resolution of the problem posed by noisy data. This is to follow Dehnen (1995) in smoothing the model between standard Richardson–Lucy corrections. Our smoothing procedure is as follows. The model is defined by the logarithms of the luminosity density at points on a $30 \times 30 \times 20$ Cartesian grid. This is aligned with the assumed principal axes of the Galaxy and covers the cuboid $0 \leq x, y \leq 5\,\text{kpc}$, $0 \leq z \leq 1.4\,\text{kpc}$ – by the Galaxy's assumed mirror symmetries, the density is then specified by the grid values throughout the region $|x|, |y| \leq 5\,\text{kpc}$, $|z| \leq 1.4\,\text{kpc}$. At each fixed value of $z$ we obtain by bi-linear interpolation the logarithm of the density on each of 35 circles whose radii are uniformly distributed over the interval $(0, 5\sqrt{2}\,\text{kpc})$; there are 60 points per circle. The routine `SMOOFT` of Press et al. (1986) is then used to smooth these data over four angular points, i.e., a smoothing angle of $24°$. Once the data for all circles in a plane of fixed $z$ have been smoothed, `SMOOFT` smooths the data along rays of fixed $z$ and azimuth $\phi$, with a smoothing length of two points. Finally the data for all circles are interpolated back onto the Cartesian grid. We do not smooth in $z$. Our smoothing algorithm has the following characteristics: (i) near the centre it does not degrade the already limited angular resolution that is provided by our Cartesian grid; (ii) the radial smoothing would not modify an exponential profile. Our results do not depend sensitively on the details of the smoothing algorithm.

## 3 RESULTS

### 3.1 Surface brightness distributions

Fig. 1 shows fits to the data obtained by deprojecting the $L$-band data for an assumed angle $\phi_0 = 20°$ between the long axis of the bar and the Sun–centre line. The full curves show contours of constant observed surface brightness after correction for extinction as in Paper II. Dotted curves show corresponding contours for the initial analytic fit to the luminosity density. This is a superposition of a double-exponential disk with a truncated power-law bulge (Paper II)

$$j(\boldsymbol{x}) = j_0 \left[ f_{\text{b}}(\boldsymbol{x}) + f_{\text{d}}(\boldsymbol{x}) \right] \tag{1a}$$

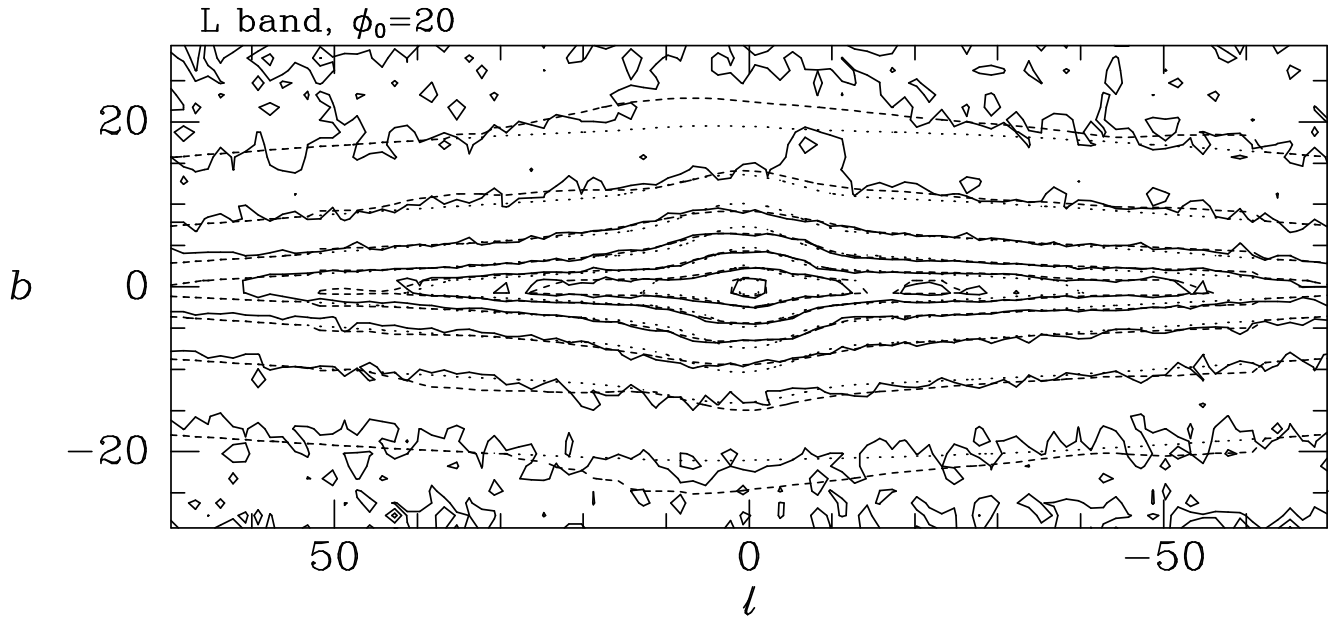

**Figure 1.** The fit between data (full contours) and model (dashed contours) that is obtained after five iterations of the Richardson–Lucy algorithm under the assumption that $\phi_0 = 20°$. The dotted contours show the initial analytic fit of equation (1). Contours are spaced by 1 mag. The Sun-centre line is assumed to lie 0.1° below the plane.

where

$$f_b \equiv f_0 \frac{e^{-a^2/a_m^2}}{(1 + a/a_0)^{1.8}}$$

$$f_d \equiv \left( \frac{e^{-|z|/z_0}}{z_0} + \alpha \frac{e^{-|z|/z_1}}{z_1} \right) R_d e^{-R/R_d} \quad (1b)$$

$$a \equiv \left( x^2 + \frac{y^2}{\eta^2} + \frac{z^2}{\zeta^2} \right)^{1/2} \quad \text{and} \quad R \equiv (x^2 + y^2)^{1/2},$$

and the constants are as follows: $j_0 = 0.456$ in COBE units for the $L$-band, $f_0 = 624$, $a_m = 1.9$ kpc, $a_0 = 100$ pc, $R_d = 2.5$ kpc, $z_0 = 210$ pc $z_1 = 42$ pc, $\alpha = 0.27$, $\eta = 0.5$, $\zeta = 0.6$.

Although the analytic model fits the data very well, *systematic* differences between the dotted contours of the analytic model and the full data contours are apparent. For example, at $l \simeq 10°$ the dotted contours fall inside the full contours on both sides of the plane. Also, near the plane the dotted contours are less elongated than the full contours.

The dashed curves in Fig. 1 show the surface brightness of the model that is obtained after five iterations of the Richardson–Lucy algorithm. (The parameter $\beta$ of Paper I was set to 0.2 for the first three iterations, and to unity for the final two iterations.) The fit to the data is now essentially perfect within the projection onto the sky of the box within which iterative adjustments are made. At $|b| \gtrsim 20°$ and $|l| \gtrsim 45°$ systematic deviations between data and model exist because much of the brightness observed in these directions derives from outside the box. In the following we largely ignore these directions and focus on the region $|l| \leq 30°$, $|b| \leq 15°$.

Five iterations of the Richardson–Lucy algorithm reduced the rms deviation between model and data within this region from 0.216 mag to 0.134 mag. The final iteration reduced the residuals by only 0.001 mag. Fig. 2 illustrates the way in which the iterations improve the fit to the data by plotting average surface brightness $I$ as a function of latitude $l$ at two values of $|b|$: $|b| < 1.5°$ (upper panel) and $3 < |b| < 4.5$ (lower panel). From the upper panel it is evident that the iterations effect a major improvement to the lowest-latitude data by fashioning a strongly non-exponential disk. At higher latitudes the iterations make smaller changes, but these include successfully modelling significant asymmetry in latitude at $|l| \lesssim 10°$. In fact, this figure shows that the final model fits the data nearly as well as any smooth model could, and that the remaining residuals are associated with small-scale structure which it is not appropriate to model at this stage.

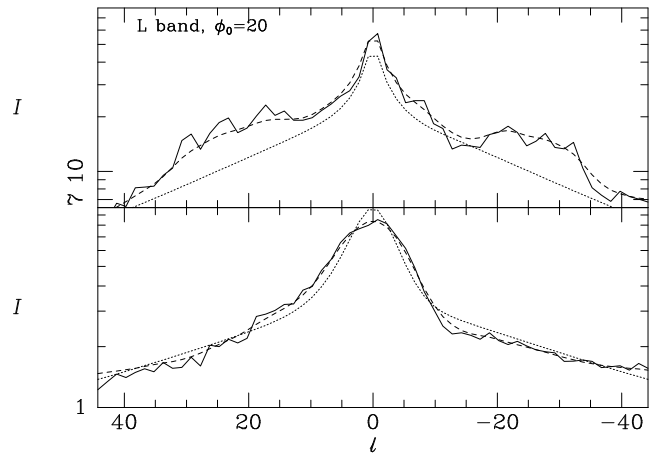

**Figure 2.** Each curve shows as a function of $l$ the average of surface brightnesses above and below the plane. The upper pane is for $|b| < 1.5°$ and the lower panel is for $3° < |b| < 4.5°$. Full curves show the data, dashed curves the final model and dotted curves the initial analytic fit.

The fit plotted in Figs 1 and 2 was obtained under the assumption that the Sun-centre line lies 0.1° below the assumed symmetry plane of the Galaxy. That is, the Sun has been assumed to lie 14 pc below the plane. Fig. 3 compares the residuals (model–data) that one obtains for this case with those that one obtains when the Sun is located within the plane (top panel) or 28 pc below the plane (bottom panel). Whereas in the top panel positive residuals tend occur at $b > 0$, in the bottom panel they occur at $b < 0$. From the fact that in the middle panel positive and negative residuals show no pronounced bias towards either positive or negative latitudes, we infer that the Sun lies of order 14 pc below the plane.

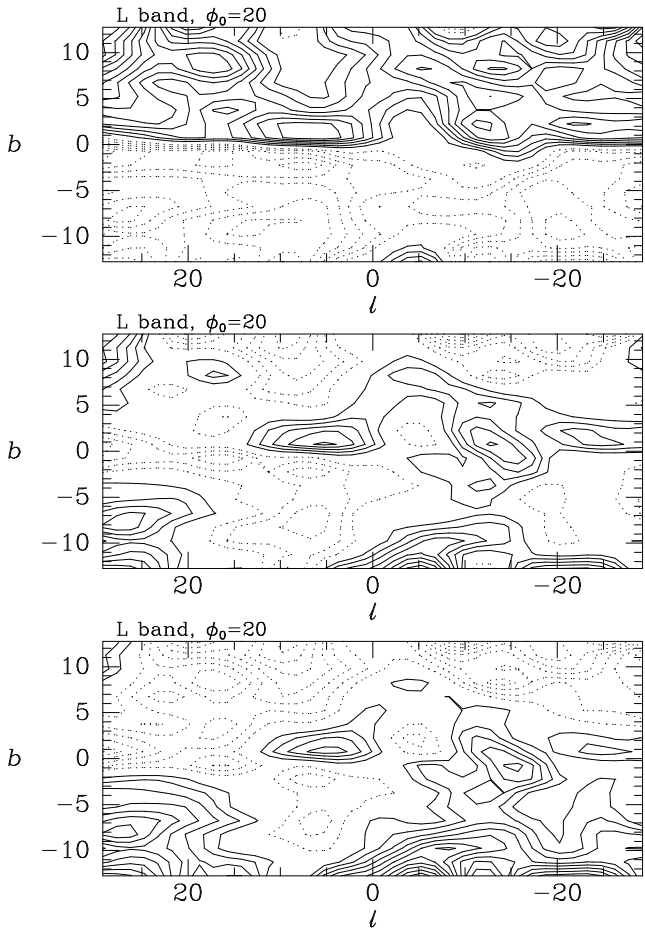

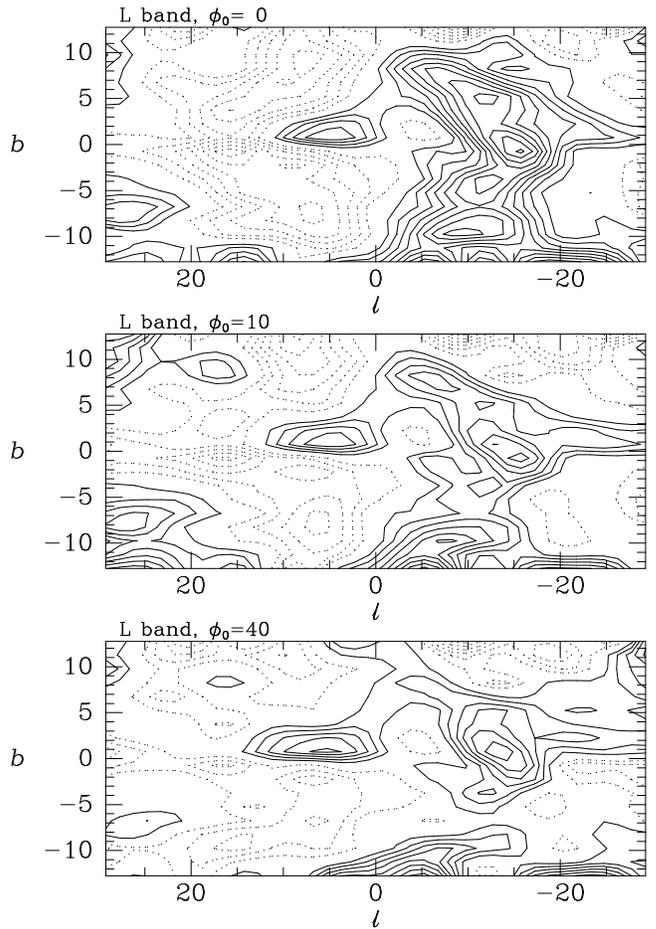

**Figure 3.** Residuals (model–data) for three different assumed locations of the Sun: in the plane (top); 14 pc below the plane (middle); 28 pc below the plane (bottom). In each case $\phi_0 = 20°$ is assumed as in Fig. 1. Contours are spaced by 0.03 mag. The Press et al. routine SMOOFT has been used to smooth the raw residuals to a resolution of 7.5° in $l$ and 3° in $b$. The middle panel is for the model of Figs 1 and 2.

**Figure 4.** Residuals (model–data) for three different values of the angle $\phi_0$ between the Sun–centre line and the assumed long axis of the bar: $\phi_0 = 0$ (top; $\phi_0 = 10°$ (middle); $\phi_0 = 40°$ (bottom). The contour levels and smoothing are as in Fig. 3. In all three panels the Sun lie 14 pc below the plane.

We now ask to what extent it is possible to constrain the angle $\phi_0$ between the Sun–centre line and an assumed principal axis by consideration of the residuals between model and measured surface brightness distributions. Fig. 4 shows the residuals (model–data) that are obtained under the assumptions $\phi_0 = 0$ (top panel) $\phi_0 = 10°$ and $\phi_0 = 40°$. At $\phi_0 = 0$ the model surface brightness distribution is symmetric in $l$ so that the preponderance of negative contours on the left of the top panel of Fig. 4 reflects the fact that the observed distribution is brighter at $l > 0$ than at $l < 0$. As Blitz &

Spergel (1991) first pointed out, this phenomenon suggests that the three-dimensional luminosity density is barred, with the nearer end of the bar lying at $l > 0$. The preponderance of negative contours at $l > 0$ is less pronounced in the middle panel of Fig. 4, which shows the residuals for $\phi_0 = 10°$, but it is still noticeable. The residuals for $\phi_0 = 40°$ shown at the bottom of Fig. 4 are remarkably similar to those for $\phi_0 = 20°$ shown in the middle panel of Fig. 3. Now positive and negative residuals are fairly impartially distributed at positive and negative $l$. For both values of $\phi_0$ there are pronounced islands of positive residuals centred on $(l, b) \simeq (7, 0)$ and $(-12, 0)$. The cause of these islands is unclear. Detailed comparison of the residual maps for $\phi_0 = 20°$ and $40°$ reveals that the cleaner map is the one for $\phi_0 = 20°$, however. In particular, the ridge of positive residuals that runs diagonally from $(l, b) \simeq (0, 12)$ to $(l, b) \simeq (15, 0)$ is distinctly less pronounced in the map for $\phi_0 = 20°$. Thus we tentatively conclude from the photometry that $\phi_0$ lies in the range $15° \lesssim \phi_0 \lesssim 35°$.

### 3.2 Luminosity densities

Fig. 5 shows the slice $z = 0$ through the model that is obtained under the assumption that $\phi_0 = 20°$ and is shown in

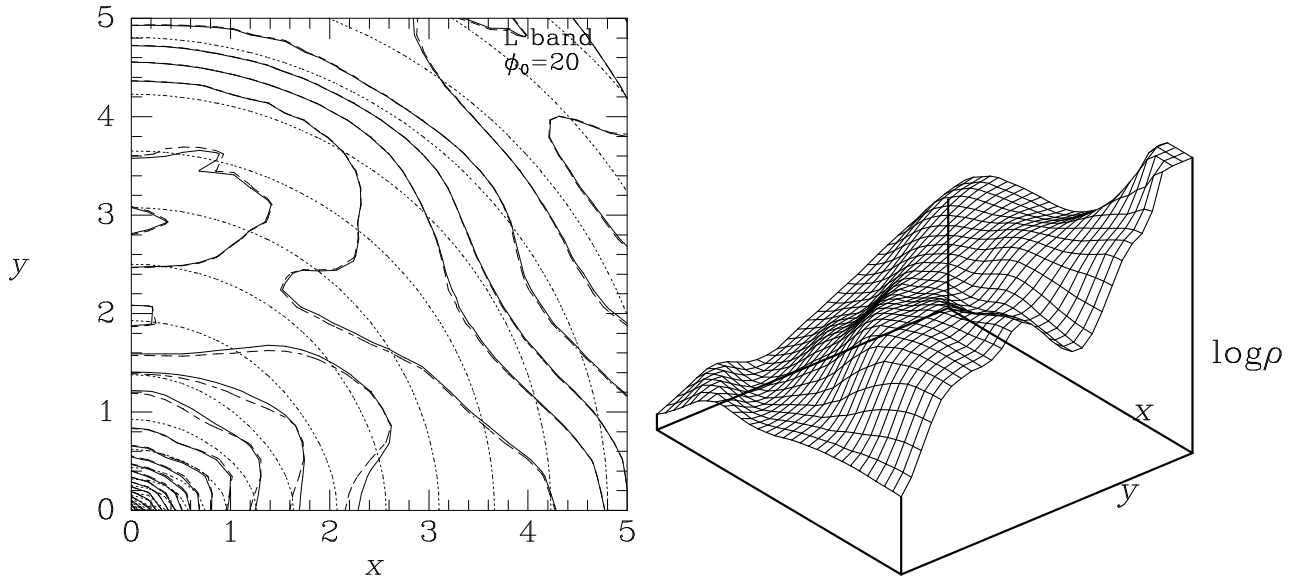

**Figure 5.** The density within the plane of the model for $\phi_0 = 20°$ whose projection is shown in Fig. 1. The three-dimensional surface at right shows the same model as the full contours at left. This model is obtained by five Richardson–Lucy iterations starting from an analytic fit which is shown by the dotted contours at left. The dashed contours show the final model that is obtained from a different initial analytic fit. All contours are separated by 0.1 dex. The central luminosity peak has been clipped in the right-hand panel. The scales on the axes in this and all similar figures are marked in kiloparsecs.

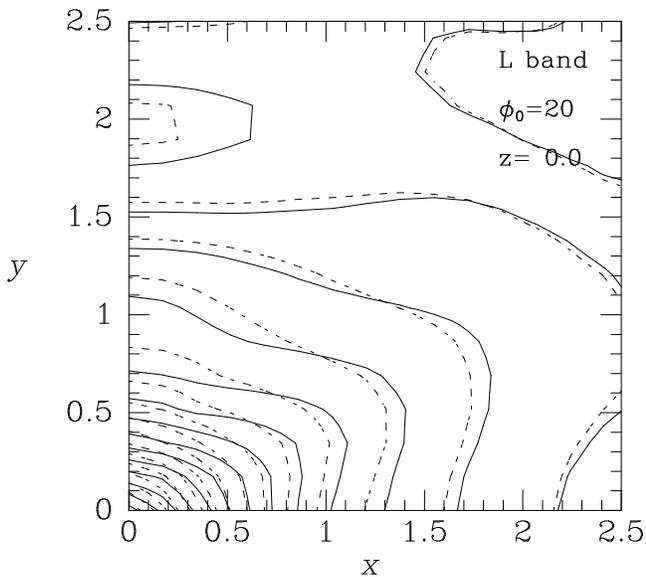

**Figure 6.** The full contours show the isodensity surfaces of the model that is obtained when the index of the central cusp's power law is changed from $-1.8$ to $-1$ [equation (1a)]. For comparison, the dashed contours are the same as those plotted in Fig. 5. This plot is analogous to Fig. 5 except that it shows only the inner 2.5 kpc of the model.

projection by Fig. 1. In the left-hand panel the full contours show the intersections with the plane of the final model's isodensity surfaces, while dotted contours show the same contours for the initial analytic fit. The right-hand panel displays the data associated with the full contours as a three-dimensional plot of $\log j$. The grid from which these plots were made has spacing 172 pc in $x$ and $y$ and 74 pc in $z$, in each case smaller than the smallest cell (210 pc) at the

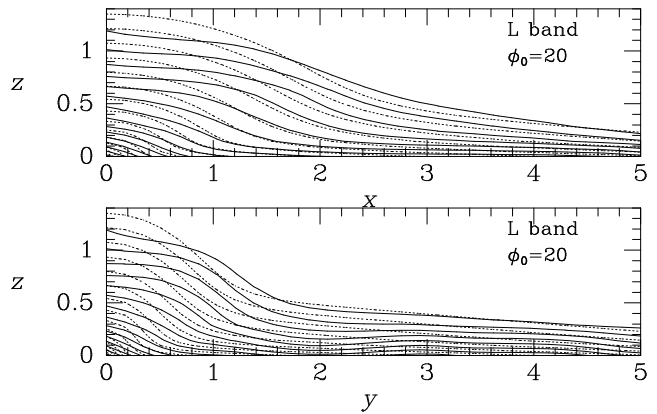

**Figure 7.** The same as the left-hand panel of Fig. 5 except for the $zx$ and $zy$ planes. Here contours are spaced at 0.2 dex, twice the spacing of the contours in Fig. 5

Galactic centre that would be resolved in the data.

The model derived by any Richardson–Lucy algorithm depends upon the model from which the iterations start. The dashed contours in Fig. 5 give a sense of the degree of this dependence in the present case by showing the model that is obtained when the parameter $\eta$ in equation (1b) is changed from 0.5 kpc to 0.8 kpc with a corresponding change in the normalizing constant $f_0$ in equation (1a) to hold the total bulge luminosity invariant. Although this change makes the initially fitted bulge less elongated in the $xy$ plane, the difference between the full and dashed contours is everywhere small. In particular, within $\sim 1.5$ kpc from the centre, where the initial models differ most, the final models scarcely differ at all.

The greatest sensitivity of the final model proves to be to the power-law index that characterizes the initially as-sumed bulge profile [equation (1a)]. For example, if $\phi_0 = 20°$ and $\alpha = 1.8$, the cusp has axis ratio $\sim 0.6$ in the plane, while for $\phi = 20°$ and $\alpha = 1$, the axis ratio is $\sim 0.55$ – see Fig. 6. Conversely, steepening the cusp from $\alpha = 1.8$ to $2.3$ makes the central isodensity surfaces significantly rounder in the $xy$ plane without significantly affecting the quality of the fit to the data that the model furnishes. This result is to be expected for two reasons. First, data with $1.5°$ angular resolution cannot resolve a central cusp. Second, to obtain a given density contrast between two points at the same galactocentric distance but down two different principal axes, one requires the isodensity contours to be less elongated the steeper the radial density profile is.

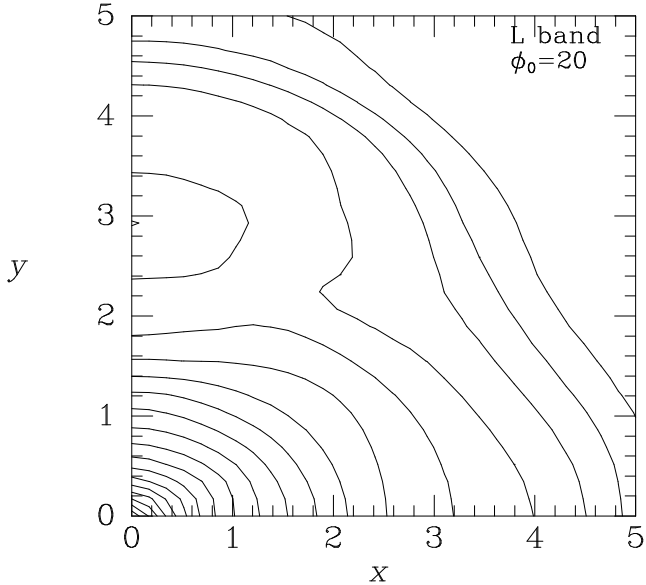

**Figure 8.** The density of the model of Figs 5 and 6 projected along the $z$-axis. Contours are spaced by 0.1 dex.

In the contour plot in Fig. 5 a bar of semi-major axis length $\sim 3.5$ kpc is evident. The crowding of the logarithmically-spaced contours towards the origin indicates that the centre of this bar has an approximately power-law density profile. The bar's outer contours deviate significantly from ellipses.

Fig. 7 shows the intersections of the $\phi_0 = 20°$ model with the $zx$ and $zy$ planes. In the lower panel the transition from the bulge to the disk manifests itself as a sharp downward section of the contours at $y \sim 1$ kpc. This transition is less evident in the upper panel, but can still be traced near $x = 1.8$ kpc. The difference between the extent of the bulge along the $x$ and $y$ axes indicates that the bar that is evident in Fig. 5 is a fully three-dimensional structure more properly associated with the bulge than the disk. Judging by the $z$-intercepts and steepest portions of the third lowest contours in Fig. 7, we deduce that the bulge has axis ratios 1:0.6:0.4. Above $z \simeq 0.4$ kpc the contours of Fig. 7 are roughly equally spaced in $z$, indicating that the bar has an approximately exponential vertical density profile away from the plane. The contours near the centre in Figs. 5 through 8 become more closely spaced as the origin is approached; this is consistent with the density profile tending to a power law (Bailey xx).

Fig. 8 shows the surface density of the $\phi_0 = 20°$ model when it is viewed along the $z$-axis. The outer contours of the bar still deviate from ellipses, but less markedly so than in Fig. 5, which shows only the equatorial plane.

A useful measure of the thickness of the Galaxy at a given point in the plane is provided by the scale height $z_0$. This is measured by least-squares fitting the recovered run of $\ln j(z)$ for $|z| \leq 295$ pc to the formula $\ln j = \ln j_0 - z/z_0$. Fig. 9 shows plots of $z_0(x, y)$ for the model depicted in Figs 5, 6 and 7. Near the centre, the scale height increases rapidly outwards from 97 pc to peak values of order 220 pc. This increase shows that the density profile of the inner bulge is a non-separable function of $R$ and $z$ and is consistent with it being of the form $j(a)$ where $a$ is defined as in equation (1b). Along the $y$-axis the peak in $z_0$ is reached at $y \simeq 600$ pc, whereas along the $x$-axis the peak is reached only at $x \simeq 1$ kpc, consistent with the power-law bulge having axis ratio $b/a \simeq 0.6$. Outside the region that is dominated by the bulge, $z_0$ falls slowly with increasing distance from the centre, reaching $z_0 \sim 120$ pc 5 kpc down the $y$ axis and $z_0 \sim 150$ pc 5 kpc down the $x$ axis. These values are considerably smaller than the value $z_0 \simeq 300$ pc at the Sun that is implied by star counts at the Galactic poles (Gilmore & Reid 1983), but entirely consistent with the findings of Kent, Dame & Fazio (1991) from Infrared Telescope data. The conflict with star counts may indicate that supergiants, which would be expected to have a small scale height because they are young, contribute significantly to the near-infrared luminosity density.

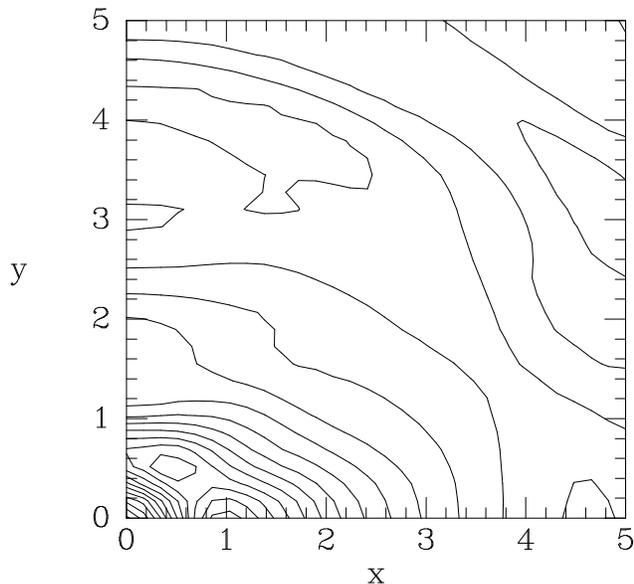
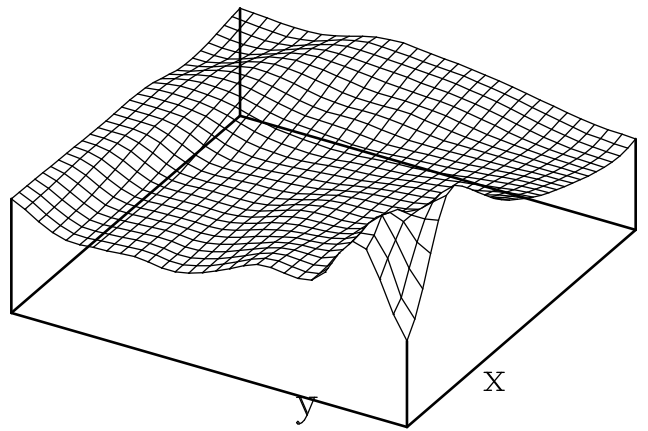

**Figure 9.** A contour plot and a three-dimensional plot of the scale-height $z_0(x, y)$ for the model shown in Figs 5 to 7. Contours are spaced by 10 pc.

Figs. 5 to 8 indicate that the bar is a fully three-dimensional object that has a power-law core out to $x \simeq 2$ kpc embedded in a roughly exponential elliptical annulus that reaches to $x \simeq 3.5$ kpc. Beyond $R \simeq 4$ kpc the density declines steeply at all azimuths as one would expect in the region immediately outside the bar. A prominent local maximum is apparent $\sim 3$ kpc down the $y$ axis. This feature proves to be strongly confined to the disk: it is absent on a plot of the density projected along $z$ for $|z| > 0.3$ kpc only. For $\phi_0 = 20°$ the points $(x, y) = (0, \pm 3$ kpc$)$ are seen along the directions $l = -22°$ and $l = 17.3°$. In Fig. 1 a subsidiary maximum is apparent at $l \simeq -22°$ in the plane and there is a hint of a local maximum at $l \simeq 17°$ in the region that is dominated by the nearer end of the bar.

Fig. 10 shows that the $M$-band data yield a model which is broadly similar to that obtained from the $L$-band data under the same assumptions regarding the orientation of the system. The residual map shown in the upper panel of Fig. 10 is also very similar to that shown for the $L$ band in the middle panel of Fig. 3.

Fig. 11 shows the model that one obtains under the assumption that $\phi_0 = 40°$. Interior to $r \simeq 1$ kpc this model is only slightly elongated in the $xy$ plane, and shows a tendency to square contours. This stubby, square bar is embedded in a long thin region of enhanced density that reaches to $x \sim 4$ kpc and $y \sim 2$ kpc, where a deep trough in the density appears. A trough on the $y$ axis is also apparent in the $\phi_0 = 20°$ model (Fig. 5), but it is significantly more pronounced in the $\phi_0 = 40°$ model.

## 4 EFFECTS OF SPIRAL STRUCTURE

Studies of external galaxies (Buta 1995) suggest that in reality the disk may be dominated by two tightly wound spiral arms that only approximately form a four-fold symmetric luminosity distribution. Moreover, both observations and $N$-body models (Sellwood & Sparke 1988, Dehnen 1996) suggest that the principal axes that are approximately defined by these arms may slightly trail those of the bulge. Any such offset between the principal axes of the bulge and the surrounding disk would be suppressed by our inversion technique. We now investigate how the existence of spiral arms would affect the models described above.

In Paper I we demonstrated the ability of the eight-fold scheme to recover true densities from noisy data, but we did not explore its performance in the presence of spiral structure. Fig. 12 shows a test of this performance. The left-hand panel shows isodensity contours within the equatorial plane of the spiral model that is obtained by multiplying the density $j_d$ of equation (1a) by the factor

$$1 + \epsilon \cos^6 \left[ \phi - k_R(r - r_{\max}) - \frac{1}{3}\pi \right], \qquad (2)$$

where $\epsilon = 2 \tanh(r - r_{\max})$ and $k_R = 0.5$ kpc. Pseudo-data were constructed by projecting this model along lines of sight to the Sun and were then deprojected with the standard Lucy–Richardson algorithm with the initial model given by equations (1). The full contours in the right-hand panel of Fig. 12 show the projection along $z$ of the resulting eight-fold symmetric model; dashed contours show the same projection of the true four-fold symmetric model. Near the centre the recovered contours agree well with the true ones. Further out, where the true contours deviate significantly from mirror symmetry within the plane, the full contours are quite different from the true ones, and there is a suggestion of an incipient trough along the $y$ axis of the type seen in Fig. 8.

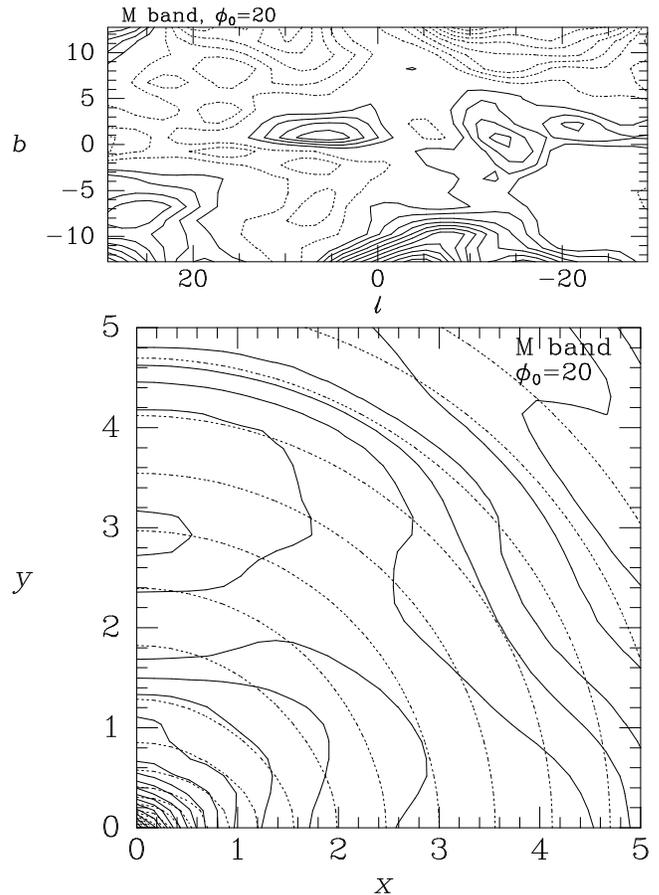

**Figure 10.** Results from modelling the $M$-band data with $\phi_0 = 20°$. The upper panel is analogous to Fig. 3, while the lower panel is analogous to the left-hand panel of Fig. 5.

Our inversion technique depends crucially on the assumption that the luminosity density is eight-fold symmetric. In the presence of spiral arms the Galaxy would be at most four-fold symmetric. It is a simple matter to reformulate the technique of Paper I for the reduced symmetry appropriate to a spiral galaxy, namely reflection symmetry with respect to the plane $z = 0$ combined with inversion symmetry within the plane. A nice feature of this algorithm is that no prior assumption need be made about the orientation of the Sun–centre line within the plane; incrementing the assumed value of $\phi_0$ by $\Delta\phi_0$ simply rotates the recovered distribution by $\Delta\phi_0$. Our tests of this algorithm all started from the initial analytic fit of equations (1) but with $\eta = 0.8$ and $f_0 = 390$. Unfortunately, in these tests this four-fold algorithm did not display the robust ability to recover true densities from their projections that was demonstrated for the eight-fold scheme in Paper I. Fig. 12 illustrates this phenomenon by showing what the four-fold scheme recovers from the projection of a four-armed spiral. The dotted con-tours are equidensity contours of the model that is obtained by multiplying the density $j_d$ of equation (1a) by the factor

$$1 + \epsilon \cos^6\left[2\phi - k_R(r - r_{\max})\right], \qquad (3)$$

where $\epsilon$ is defined below equation (2) and $k_R = 1$ kpc. The full contours show the eighth iterate of the four-fold algorithm. Of the four arms only two appear in the reconstruction, and these are significantly out of focus. The recovered arms are those that are seen approximately end-on and thus give rise to significant enhancements in the surface-brightness pseudo-data. Conversely, the two missing arms are those that seen broad-on. Thus the result of using the four-fold algorithm to deproject the COBE data will depend on exactly how the Galaxy's spiral arms are oriented with respect to the Sun–centre line.

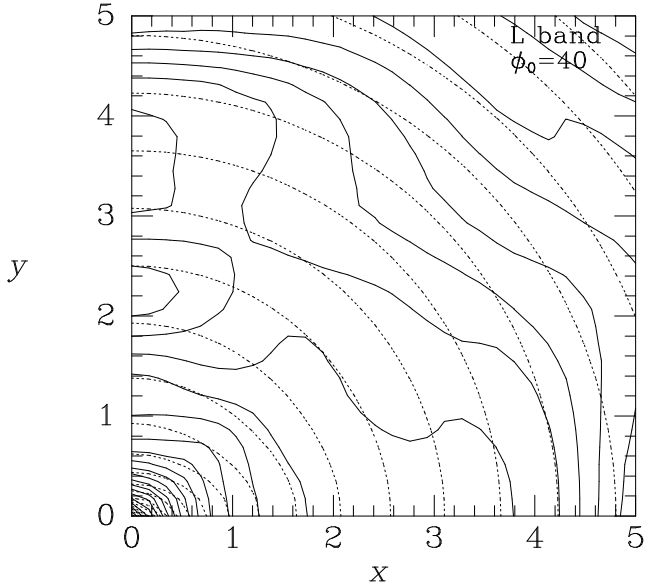

**Figure 11.** The density within the plane of the model obtained from the $L$-band data under the assumption that $\phi_0 = 40°$.

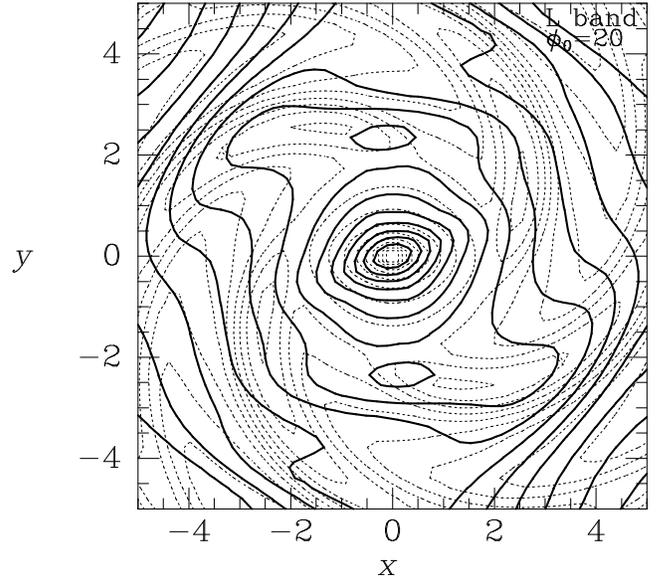

**Figure 13.** The dotted contours show isodensity contours in the equatorial plane of a model that includes four-armed spiral structure. The full contours show the result of using the four-fold Lucy–Richardson algorithm to deproject an image of this model. The direction to the Sun makes an angle of $20°$ with the $x$ axis but no assumption about the orientation of principal axes in the plane has been made.

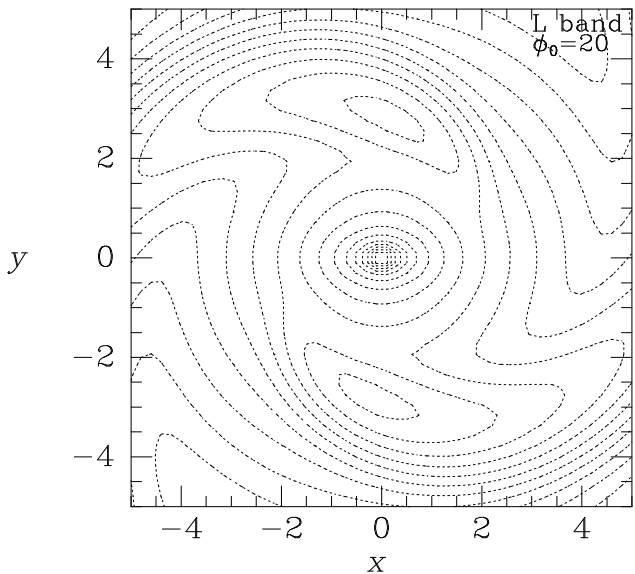

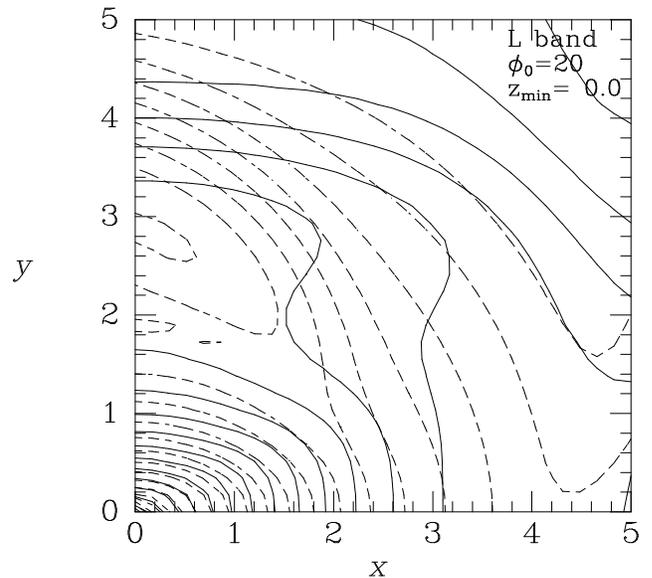

**Figure 12.** A test of the ability of the eight-fold technique to cope with spiral structure. The spiral model shown in the left panel was projected along lines of sight from the Sun and then deprojected under the assumption of eight-fold symmetry. The full contours at right show the projection along $z$ of the deprojected model, while the dashed contours show the projection of the original model.

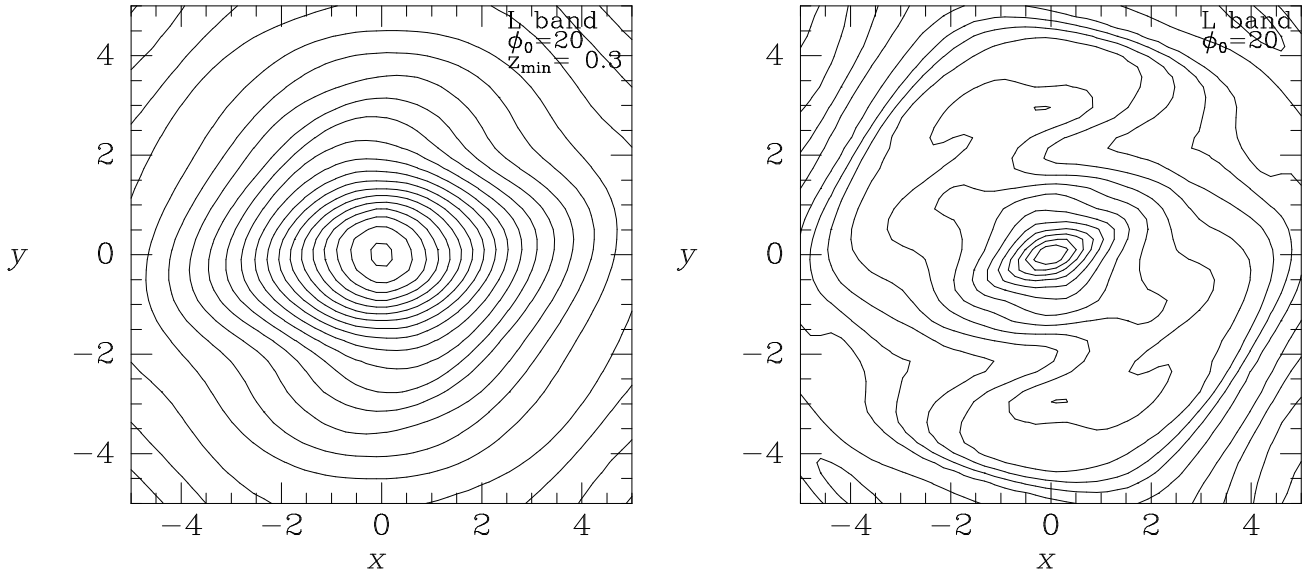

**Figure 14.** The model obtained by deprojecting the COBE $L$-band data with the four-fold algorithm. The left panel shows the projection parallel to $z$ of the luminosity at $|z| > 300\,\mathrm{pc}$. The right panel shows the density in the plane $z = 0$. The initial analytic model from which the iterations were started was axisymmetric.

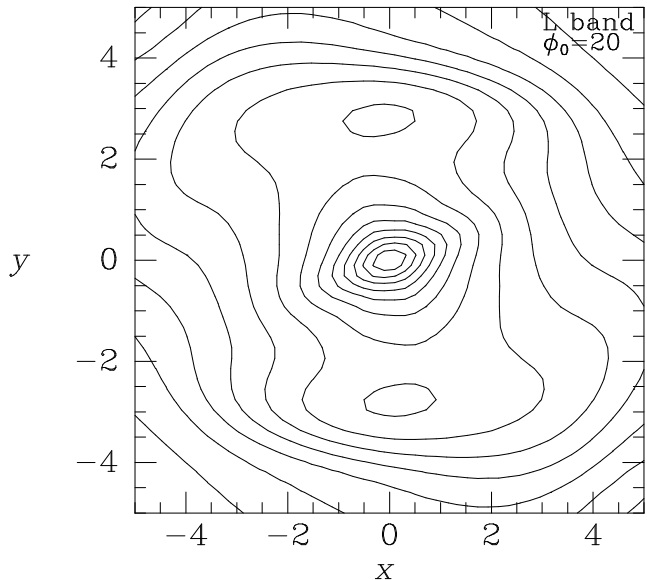

**Figure 15.** The result of deprojecting an image of the spiral model that is shown in the left panel of Fig. 12.

Fig. 14 shows the result of applying the four-fold algorithm to the COBE data. The left-hand panel shows the projection parallel to $z$ of all the luminosity that lies further than 300 pc from the plane, while the right-hand panel shows the slice $z = 0$ through the model. (300 pc is the smallest distance from the plane at which the model along $x = 0$ is uninfluenced by the data for $b = 0$.) Two things are noteworthy about this figure.

(i) The left panel, which is presumably dominated by bulge luminosity, shows that the model is broadly similar to that obtained under the assumption of eight-fold symmetry, especially near the centre. The long axis of the recovered luminosity distribution points approximately along the $x$-axis. In the algorithm used here, the orientation of the coordinate axes is of no special significance, whereas relative angles between the Sun–centre line and the principal axes of the recovered luminosity distribution are significant. Thus the left panel in Fig. 13 strongly implies that the major axis of the bar makes an angle close to $20°$ with the Sun–centre line. To eliminate any possibility of bias, the inversion shown in Fig. 13 was started from an axisymmetric initial model, although the model obtained with the initial condition that was used for Figs. 13 and 15 is essentially identical.

(ii) In the equatorial slice shown in the right-hand panel of Figure 14, the region around the bar is strongly non-axisymmetric. In particular, it displays local maxima at $\sim 3\,\mathrm{kpc}$ along the $y$ axis that are very similar to the local maxima on the $y$ axis of the eight-fold symmetric models. Enhanced emission along the $y$ axis can be seen in slices through the model at $|z| > 0$ out to $|z| = 440$ pc, although the strength of this emission decreases rapidly with $|z|$. The fact that this emission peaks on the $y$ axis is significant because there is nothing inherently special about the $y$ axis in the four-fold scheme. So there really must be something in the data that requires these features to lie along the $y$ axis. In the equatorial slice of Fig. 14, the inner isodensity contours are twisted by about $15°$ towards the Sun–centre line. Since this effect does not persist to high $z$, we suspect that it is an artifact associated with the presence of misaligned, perhaps spiral, structure in the Galactic disk.

An odd feature in the right panel of Fig. 14 is that the maxima on the minor axis appear to be parts of *leading* spiral arms – in the figure the Galaxy rotates clockwise. This leading form is definitely required by the data in as much as when one projects the model that is obtained by reflecting the picture in the $y$ axis ($x \to -x$), the residuals between model and data increase significantly. In Fig. 12 the deprojected arms wind in the same sense as the true arms. Thus Fig. 14 suggests that the inner Galaxy possesses leading spiral arms.

In view of this situation we have experimented with deprojecting the images of spiral patterns that we declare to be trailing to see whether leading deprojected arms can be obtained for some orientation of the Sun relative to the true arms. Fig. 15 shows our best effort – this model is obtained by deprojecting an image of the model that is shown in the left panel of Fig. 12. The model of Fig. 15 is crudely similar to the COBE deprojection of Fig. 14 in that (i) there are local maxima on the minor axis, and (ii) there is structure which might be interpreted as leading arms. However, Fig. 15 shows trailing structure, which, although much weaker than in the true distribution shown in Fig. 12, is still much stronger than any trailing structure in the deprojected COBE image (Fig. 14).

Three further points are noteworthy about Fig. 15. First, only *strong* spiral arms have a significant effect on the deprojected image – the arms shown in the left-hand panel of Fig. 12 have peak amplitude of 300%. Second, the local maxima on the minor axis of the deprojected distribution shown in Fig. 15 are not artifacts, but are pale reflections of stronger maxima in the underlying spiral arms shown in the left-hand panel of Fig. 12. Finally, the long axis of the recovered bar is twisted by about $15°$ towards the Sun–centre line. In the right-hand panel of Fig. 14 the central contours in the COBE deprojection are twisted by a very similar amount in the same direction. This confirms our suspicion that this twist in the COBE reconstruction is an artifact introduced by spiral structure in the Galactic plane.

We draw the following conclusions from these experiments.

(i) The long axis of the bulge must make an angle with the Sun–centre line that lies close to $20°$.

(ii) Spiral arms of large amplitude and appropriate orientation can produce features within the plane like those found in the models of the Galaxy that are recovered by the eight-fold and the four-fold algorithms.

(iii) Our tests suggest that spiral structure can account for the local maxima evident on the $y$ axes of our models only if the spiral arms have local maxima at about the same locations. Therefore these local maxima are likely to be real.

(iv) Spiral structure cannot be reliably recovered by Lucy–Richardson inversion. Nonetheless, in the great majority of tests of the four-fold algorithm, the sense of the input model's spiral arms was correctly reproduced in the deprojected model. Thus it is worrying that the deprojected model of the Milky Way shown in Fig. 14 appears to contain leading spiral arms.

## 5  CONCLUSIONS

We have obtained three-dimensional models of the luminosity density of the Milky Way interior to 5 kpc by deprojecting COBE/DIRBE near-infrared data that have been corrected for the effects of obscuration by dust. The models recovered by two different Lucy-Richardson algorithms all show a well-defined three-dimensional bar set within a highly non-axisymmetric and non-exponential disk.

Our first deprojection algorithm assumes that the Galaxy is eight-fold symmetric and requires the prior specification of the orientation of its principal axes. From a study of the maps of residuals between the observed surface brightnesses and those predicted from our eight-fold symmetric models we have concluded (i) that the angle $\phi_0$ between the long axis of the bulge and the Sun–centre line lies in the range $15° \lesssim \phi_0 \lesssim 35°$, and (ii) that the Sun lies $(14 \pm 4)$ pc below the plane. The results obtained with the $L$- and $M$-band data are remarkably similar.

Our second algorithm assumes four-fold symmetry and allows for the inclusion of spiral structure. It makes no prior assumptions about the orientation of structure in the Galactic plane. When we deproject the COBE data under this weaker symmetry assumption, we recover a model that is gratifyingly similar to that obtained when we deproject under the assumption of full eight-fold symmetry with $\phi_0 = 20°$. This result strongly suggest that $\phi_0$ lies near $20°$.

The models we recover do not depend sensitively on the model from which the Lucy–Richardson iterations start. The strongest dependence of the final model upon the initial guess is to the strength of the initial cusp. Since the data have an angular resolution of only $1.5°$, they have difficulty distinguishing strongly and weakly cusped initial profiles, and we find, as expected, that more strongly cusped initial profiles give rise to less elongated central density contours. For a given slope of the central density profile, we find that the shapes of the isodensity contours within the plane are remarkably independent of the shape of the initial model.

In our favoured models a fully three-dimensional bar extends to $\sim 1$ kpc down the minor axis and $\sim 1.8$ kpc down the major axis, with approximate axis ratios 10:6:4. Its radial density profile approximates a power law $j \sim a^{-\alpha}$ near the centre but to some extent this must reflect the assumed initial model. The vertical density profile above 400 pc is approximately exponential.

In the eight-fold symmetric deprojections, the elongation of the bar varies with $\phi_0$ in the expected sense: larger values of $\phi_0$ give rise to less elongated bulges. The three-dimensional barred bulge is surrounded by a thin elliptical annulus that extends to $\sim 2$ kpc and $\sim 3.5$ kpc on the minor and major axes, respectively, approximately independently of the assumed value of $\phi_0$.

All our models display a density trough at $y \simeq 2.2$ kpc followed by a local maximum, and by highly non-axisymmetric isodensity contours further out. These features lie within $\sim 300$ pc of the plane. Experiments with the four-fold algorithm suggest that the trough and subsidiary maximum along the $y$ axis are real and not artifacts of our deprojection technique. By contrast, the detailed shapes of the isodensity contours may well be artifacts that arise during the inversion of surface brightness distributions of inherently spiral disks. Arm-interarm contrasts exceeding 300% are required – contrasts of this order have been reported for M100 by Gnedin, Goodman & Rhoads (1996).

How should we interpret our models dynamically? We concentrate on the model for $\phi_0 = 20°$ both for the reason given above and because studies of the kinematics of gas at the galactic centre (Binney et al. 1991; Binney 1994; Gerhard 1996), microlensing in Baade's window (Alcock et al. 1995; Udalski et al. 1994; Zhao, Rich & Spergel 1995), and the distribution of red clump stars (Stanek et al.1996) strongly favour small values of $\phi_0$.

The subsidiary peak along the $y$ axis may be formed by stars trapped by the $L_4$ and $L_5$ Lagrange points – these stars are probably young and may be supergiants (see Rhoads 1996). Hence any subsidiary peak along the $y$ axis in the *mass* distribution may be smaller than our luminosity distributions suggest. In any event, if the subsidiary peaks do arise from the Lagrange points, corotation will lie at $\sim 3$ kpc. $N$-body models indicate that bars extend to $\sim 0.8$ of the corotation radius (Sellwood & Sparke 1988), so if the bar extends to 2 kpc, corotation lies at $\sim 2.5$ kpc. If we assume that the Galaxy's underlying circular-speed rises as $v_c \propto R^{0.1}$ (Binney et al. 1991) and that $v_c(R_0) = 200$ km s$^{-1}$, then from the subsidiary peak on the minor axis we conclude that the bar's pattern speed is $\Omega_b \simeq 60$ km s$^{-1}$ kpc$^{-1}$, while from the length $\sim 2$ kpc of the vertically thick bar we derive $\Omega_b \simeq 71$ km s$^{-1}$ kpc$^{-1}$. For comparison, Binney et al. inferred $\Omega_b = 63$ km s$^{-1}$ kpc$^{-1}$ from CO and HI observations interior to $l = 10°$, while Kalnajs (1996) derives $\Omega_b = 39$ km s$^{-1}$ kpc$^{-1}$ from observations of stellar kinematics in Baade's window.

The non-axisymmetric structure in our models depends upon systematic asymmetries with longitude in the brightness distribution that are $\lesssim 0.4$ magnitudes in size. Consequently, the reliability of our models hinges upon the corrections for dust obscuration made by Spergel et al. (1996) being accurate to better than $\sim 0.4$ magnitudes. These corrections are largest at the lowest latitudes, so the greatest uncertainties in our models must attach to the predicted disk structure. Our predictions for the structure of the bulge may be regarded as more robust.

It is likely that better estimates of dust obscuration could be obtained by combining the technique of Spergel et al. with that of Kent et al. (1991), who used a kinematic model of the gas distribution to localize the dust. Traditional kinematic models have the weakness that they assume circular gas motions, which is inconsistent with the presence of the Galactic bar. Given the mass models derived here, one can now derive a proper dynamical gas model, check it with the observed radio frequency emission-line distribution, and from it obtain a significantly improved gas and dust model. Given the importance of understanding the distribution of light and mass in the Galactic inner disk, this task appears well worthwhile.